\documentclass[12pt,preprint]{aastex}
\newcommand{\onee}   {{1E~2259+586}} 
\newcommand{\gone}   {{G109.1-1.0}} 
\newcommand{\chan}   {{\sl Chandra}} 
\newcommand{\beppo}  {{\sl BeppoSAX}} 
\newcommand{\bbxrt}  {{\sl BBXRT}} 
\newcommand{\asca}   {{\sl ASCA}} 
\newcommand{\rosat}  {{\sl ROSAT}}

\newcommand{\cxo}    {{\sl Chandra X-ray Observatory}} 
\newcommand{\nh}     {$N_{\rm H}$} 
 
\newcommand{\kt}     {$kT_{\rm BB}$} 
\newcommand{\msun}   {M_{\odot}} 
 
\begin{document} 
\title{\chan\ Observations of the Anomalous \\ X-ray Pulsar 
\onee} 
\author{ 
Sandeep~K.~Patel\altaffilmark{1}, 
Chryssa~Kouveliotou\altaffilmark{1,2}, 
Peter~M.~Woods\altaffilmark{1}, 
Allyn~F.~Tennant\altaffilmark{2}, 
Martin~C.~Weisskopf\altaffilmark{2}, 
Mark~H.~Finger\altaffilmark{1}, 
Ersin~{G\"o\u{g}\"u\c{s}}\altaffilmark{3}, 
Michiel~van~der~Klis\altaffilmark{4}, 
Tomaso~Belloni\altaffilmark{5} 
} 

\altaffiltext{1} {Universities Space Research Association/NSSTC, SD-50, 
Huntsville, AL 35805} 
\altaffiltext{2} {NASA Marshall Space Flight Center, Huntsville, SD-50, 
AL 35812} 
\altaffiltext{3} {Department of Physics, University of Alabama in 
Huntsville/NSSTC-SD50, Huntsville, AL 35805} 
\altaffiltext{4} {Astronomical Institute "Anton Pannekoek," University 
of Amsterdam, and Center for High Energy Astrophysics, Kruislaan 403, 
1098 SJ Amsterdam, Netherlands} 
\altaffiltext{5} {Osservatorio Astronomico di Brera, Via Bianchi 46, 
23807 Merate (Lc), Italy} 

\begin{abstract} 

We present X-ray imaging, timing, and phase resolved spectroscopy of 
the anomalous X-ray pulsar \onee\ using the \cxo. The spectrum is well 
described by a power law plus blackbody model with $\Gamma = 3.6(1)$, 
\kt$=0.412(6)$~keV, and \nh$=0.93(3)~\times~10^{22}~{\rm cm}^{-2}$; 
we find no evidence for spectral features ($0.5-7.0$ keV). We derive a 
new, precise X-ray position for the source and determine its spin 
period, $P=6.978977(24)$~s. Time resolved X-ray spectra show no 
significant variation as a function of pulse phase. We have detected 
excess emission beyond $4\arcsec$ from the central source extending to 
beyond $100\arcsec$, due to the supernova remnant and possibly dust 
scattering from the interstellar medium. 

\end{abstract} 
\section{Introduction} 
\label{sec:intro} 

X-ray pulsars, typically, have hard spectra and a broad range of 
observed spin periods (2~ms -- 3~hr). There is, however, a group of at 
least 5 X-ray pulsars (4U~0142$+$61,   1E~1048.1$-$5937, 
1RXS~J170849$-$40091, 1E~1841$-$045, 1E~2259$+$586) that   have 
characteristics similar to one another, but distinct from those known 
to be accreting binary pulsars. The existence of this class of 
pulsars, dubbed the `Anomalous X-ray   Pulsars' (AXPs), was first 
identified as such 6 years ago \citep{mereghetti1995,vanparadijs1995}. 
AXP characteristics include: (i) predominantly constant X-ray 
luminosities of order $\sim$ 10$^{35}$ ergs s$^{-1}$; (ii) relatively 
soft, two-component X-ray spectra (blackbody $kT_{BB} \sim$ 0.5 keV; 
power-law photon indices $\sim$~2.5$-$4); (iii) small Galactic scale 
height ($z_{\rm rms} \sim$ 56 pc); (iv) association of 2 out of 5 
(1E~1841$-$045, 1E~2259$+$586) with young ($\sim$ 10$^{4-5}$ year) 
supernova remnants (SNRs) \citep{gaensler2001}; (v) pulse periods 
within a narrow range (5 -- 12 s); (vi) rapid, nearly constant 
spin-down rates ($10^{-11} - 10^{-13}$ s s$^{-1}$); and (vii) no clear 
evidence for a binary companion filling its Roche lobe \citep[see e.g.][ for 
a review]{mereghetti1999}. 

\onee\ was discovered by Gregory \& Fahlman (1980) and lies along the 
line of sight of an X-ray and radio bright galactic supernova remnant 
\gone\ (CTB 109) \citep{rho1997,hughes1984}.  There is no accurate 
distance measurement for the SNR; \cite{sofue1983} and 
\cite{hughes1984} have estimated distances of 4.1 and 5.6~kpc, 
respectively, using the surface brightness-diameter relation in 
different wave bands. The X-ray properties of \onee\ (and the 
associated SNR \gone) have been investigated using \beppo, \rosat, 
\bbxrt, and \asca\ \citep{parmar1998,rho1997,corbet1995}.  The pulsar 
spectrum has often been described as an absorbed powerlaw (PL) with a photon 
spectral index of $\Gamma=4.0$ and a blackbody (BB) component with a
characteristic temperature of \kt$=0.43$~keV with galactic absorption 
column density of \nh=$0.9 \times 10^{22}$~cm$^{-2}$ \citep{rho1997}. 

The best previous position for \onee\ was $\alpha=23^{\rm h}~01^{\rm 
m}~08^{s}.44$, $\delta=58\arcdeg~52\arcmin~44\arcsec.1$ (J2000) with a 
$95\%$ confidence error radius of $2\arcsec.2$ established by 
\cite{hulleman2000a} using \rosat\ HRI observations. They found no 
optical counterpart with the Keck, down to limiting magnitudes of 
$R=25.7$ and $I=24.3$. \cite{coe1994} set an upper limit with the VLA 
to the 1.5~GHz flux of $50~\mu{\rm Jy}$. 

We present here an analysis of our \chan\ observation of the source. 
In \S \ref{sec:obsresults} we describe the observations and present 
imaging, timing and spectral results.   The analysis and derivation of 
a new precise position of $\alpha=23^{h}~01^{m}~08^{s}.295$, 
$\delta=+58\arcdeg~52\arcmin~44\arcsec .45$ (J2000) with a $99\%$ 
confidence error radius of $0\arcsec.60$, together with the K-band identification of \onee\, is discussed in the companion paper 
by \cite{hulleman2001}. 

\section{Observations and Results} 
\label{sec:obsresults} 
\onee\ was observed on 11 January 2000 using the Advanced CCD Imaging 
Spectrometer (ACIS).  Data were collected sequentially in two different 
observing modes: timed exposure (TE) mode (19 ks) and continuous clocking 
(CC) mode (12 ks). 
Data obtained in TE mode allow for two-dimensional imaging. 
Accurate spectroscopy of the bright pulsating target is limited due to 
pulse pile up. 
Pulse-phased spectroscopy is further limited by the 3.24~s time resolution.  
In CC mode the amount of pileup is negligible due to its 2.85  
ms time resolution, allowing for both accurate time integrated and 
phase-resolved spectroscopy.  
Furthermore, one can exploit the one-dimensional image to search for 
extension of the central source.

\begin{figure*}[r] 
\epsscale{1.0} 
\centerline{\includegraphics[width=6.0in, angle =-90]{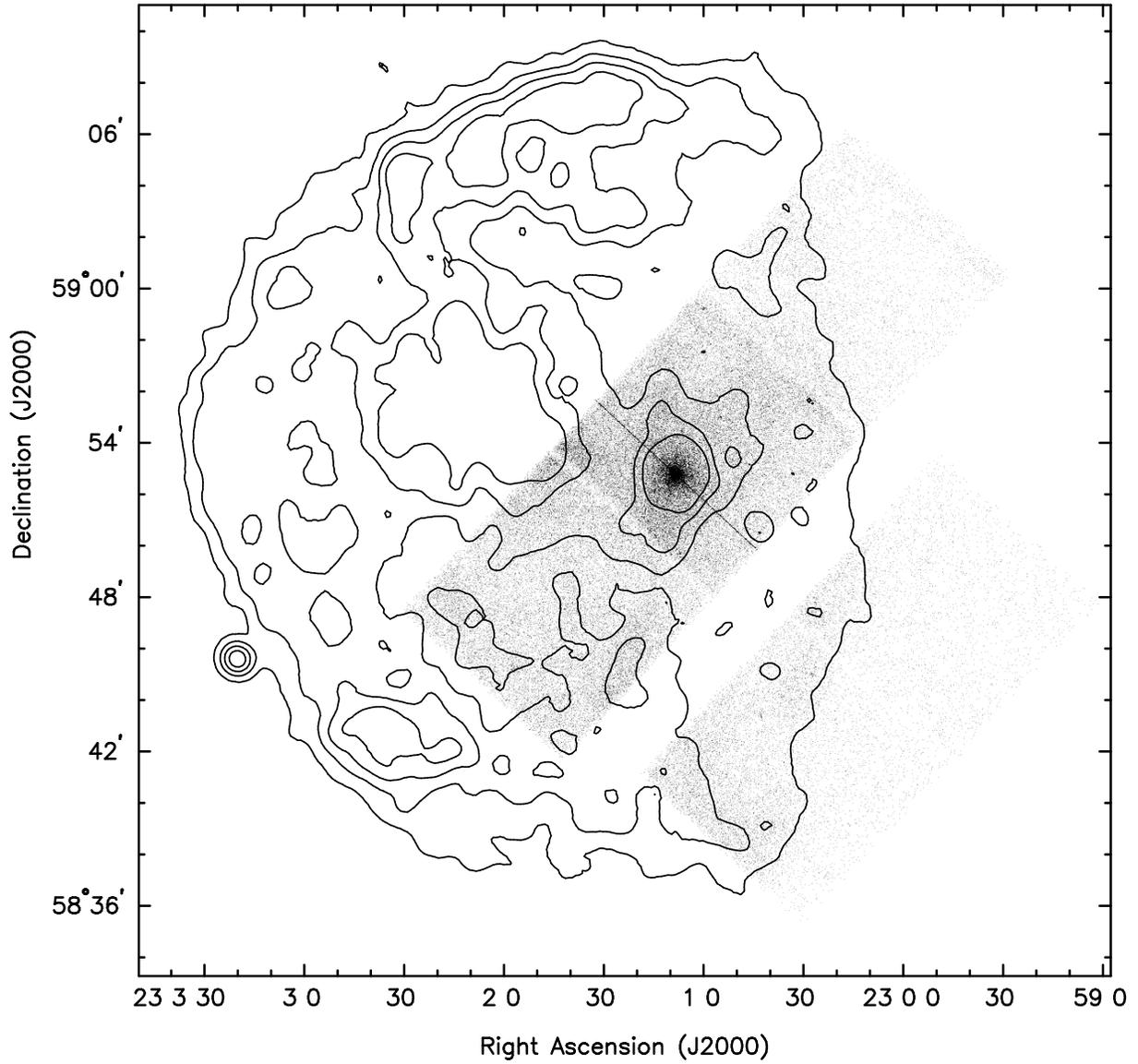}}
\caption{\chan\ ACIS $0.5-7.0$~keV image processed as described in 
\S~\ref{sec:axpspec}.  The bright line that passes throught the source is the ACIS 
transfer (or trailed) image \citep{weisskopf2000}.  
The false grey scale represents the number of counts detected. The \rosat\ image 
\citep{rho1997} is superimposed for comparison and to give a better view of the SNR. 
} 
\label{fig:chanimage} 
\end{figure*}

The source was positioned on the nominal target position of ACIS-S3, a 
back illuminated CCD on the spectroscopic array (ACIS-S) with good 
charge transfer efficiency and spectral resolution 
\citep{townsley2000b}. 
In addition to S3, four front illuminated CCDs 
were active (I2, I3, S2 and S4, see also Figure~\ref{fig:chanimage}). The 
focal plane operating temperature was $-110\arcdeg$~C. 

Standard processing of the data was performed by the \chan\ 
X-ray Center. 
The data were filtered to exclude events with \asca\ grades 1, 5, and 7,  
hot pixels, bad columns, and events on CCD node boundaries. 
We removed the serial clocking noise streaks from S4 using the {\sl  
destreak} program. 
We examined the processed data and found no times with bad aspect. 
The S3 light curve was inspected in a region offset from the AXP to 
identify periods of high background rates: we removed segments where the 
background exceeded a $3\sigma$ threshold about the mean.
The resulting useful observing times are 15600~s and 9000~s for TE and CC  
mode data, respectively. 

\subsection{Extended Emission Search} 
\label{sec:extended} 
Searching for extended emission in the vicinity of \onee\ was 
difficult. One must account for numerous contributions to the observed 
flux. These contributions include not only the bright source itself 
and the instrumental background, but also the SNR, stars 
in the field, and diffuse emission from the galactic plane. One must 
also account for the possibility of a dust scattering halo and/or an 
X-ray nebula. 

To avoid issues related to the effects of pileup in the TE mode data, 
we first utilized the one dimensional image from the CC mode. We 
generated a time-integrated image ($0.5-7.0$ keV) minus any point 
sources other than \onee. We then subtracted an average count rate 
intended to remove instrumental and cosmic diffuse X-ray background, 
as measured by S3 \citep[][and references therein]{markevitch2001}. 
Next, we constructed what will be referred to as the ``pulsed'' image. 
We took the observed pulse profile, normalized it to a mean of zero 
and convolved it with the event list of S3.  Each count 
recorded on S3 was assigned a phase. Events with a phase near pulse 
maximum, regardless of their position on the chip, received a positive 
weight and likewise, those near pulse minimum, a negative weight.  In 
doing so, we remove all emission components in our image that do not 
vary in phase (i.e., everything except the central pulsar).  Note that 
the time delay for photons scattered by the interstellar dust is on 
average minutes, much longer than the pulsar period. 

\begin{figure*}[r] 
\epsscale{1.0} 
\plotone{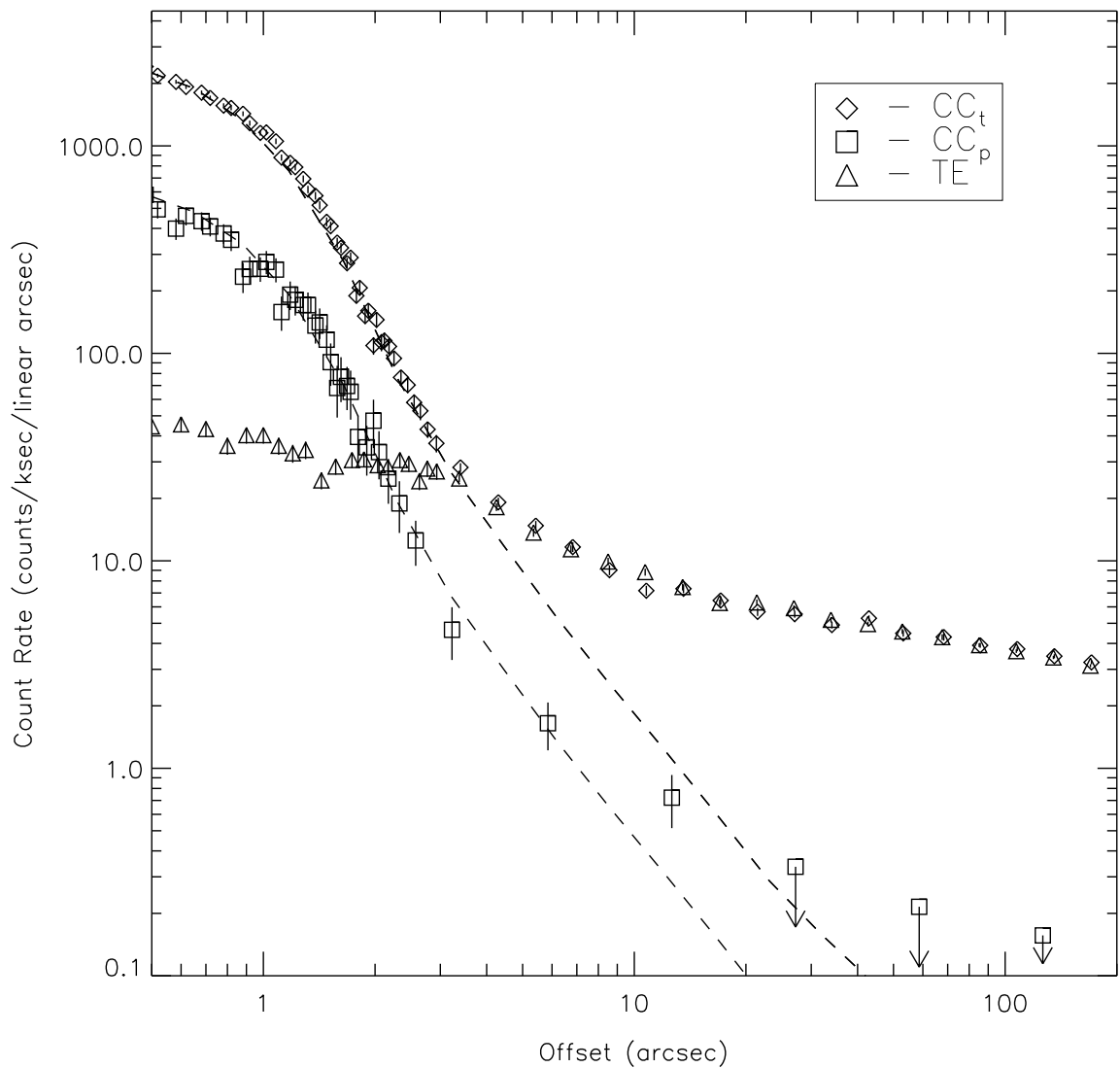}
\caption{ 
Radial surface brightness profiles of the total (CC$_{\rm t}$) and pulsed 
(CC$_{\rm p}$) emission (0.5$-$7.0 keV).  The 
dashed lines are the simulated \chan\ point spread function. 
Downward pointing arrows denote 2$\sigma$ upper limits to the count 
rate. The triangles are  TE mode data. 
Note the agreement between the total CC profile and the TE profile 
at radii $\gtrsim4\arcsec$.} 
\label{fig:radprof} 
\end{figure*}

To improve statistics, we folded the one-dimensional images about 
the common centroid and accumulated ``quasi-radial'' profiles.  The 
time-integrated (CC$_{\rm t}$) and pulsed (CC$_{\rm p}$) profiles are shown 
in Figure~\ref{fig:radprof}. 
We find that the pulsed radial profile is consistent with the MARX  
\citep[v3.01]{wise2000} derived point spread function (PSF). 

Next, we collapsed the TE mode observation to mimic the 1-D CC image 
(Figure~\ref{fig:radprof}, TE). We find that the TE profile completely 
overlaps the CC$_{\rm t}$ profile beyond $\sim 4\arcsec$, marking the 
radius beyond which the effects of pile-up are negligible. In both 
profiles we find excess emission beyond $\sim 4\arcsec$ which is  
likely due to the combination of the underlying SNR, a dust-scattering 
halo and potentially an X-ray plerion. We conclude that the majority 
of the \onee\ X-ray flux is contained within a radius of $\sim 
4\arcsec$. Inspection of the spectra  extracted from S3 in five 
annuli, centered on the pulsar position, and with inner and outer 
radii of $4\arcsec-20\arcsec$, indicates the presence of continuum 
radiation in the vicinity of the pulsar. Detailed modeling of the 
angular and spectral distribution is underway for an accurate 
accounting of the light, to establish the relative contributions from 
the SNR, dust scattered light from the pulsar, and any possible 
plerionic emission. This work will be   presented elsewhere. 

\subsection{Timing Results} 
\label{sec:axptim} 

High resolution timing with ACIS is only possible using CC-mode  
data. The CC mode event times denote when the event was read out of 
the frame store, not when it was detected. We corrected for this 
effect by assuming that all photons were originally detected at the 
nominal target position.  We removed the variable time delay due to 
spacecraft dither and telescope flexure using the ACIS CC mode absolute 
time corrector\footnote{http://wwwastro.msfc.nasa.gov/xray/ACIS/cctime/}. 
The event arrival times were then corrected to the solar system barycenter 
using {\sl axbary} which utilizes the JPL planetary ephemeris DE-200. 

The data were divided into ten $\sim1200$~s intervals and individual 
pulse profiles were derived using epoch folding.  These ten profiles 
were compared to the pulse profile using all the data and the relative 
phases were fit with a linear function.  The resulting pulse period of 
$6.978977(24)$~s is referred to epoch MJD 51555.0.  This period is 
consistent with the spin history of the source 
\citep{gavriil2001}. The pulse profile is shown in 
Figure~\ref{fig:foldevents}a; the 0.5$-$7.0 keV background subtracted 
peak-to-peak pulse fraction, as defined in \cite{ozel2001}, 
is $35.8\pm1.4\%$. 

Previously, \cite{mereghetti1998} have established limits as to the 
existence of a binary companion.  They concluded that a putative 
companion must be a white dwarf or a He-burning star with $M\lesssim 
0.8\msun$ (for a companion star filling its Roche lobe). We have also searched for evidence of binarity. Binning the 
data in 50 s intervals, a total of 177 pulse profiles were 
cross-correlated to the average pulse profile, and 
pulse phase offsets were determined. We then searched for the 
sinusoidal signature of a circular binary orbit, for periods 
ranging from 100 to $5\times 10^4$~s using the method described by 
\cite{wilson1999}. We found no evidence for binarity and set a $99\%$ 
confidence limit of $a_x \ {\rm sin}~i < 70$~lt-ms for orbital 
periods in the range from  170 to 5000~s \citep[see also ][]{mereghetti1998}. 

\subsection{Spectral Analysis} 
\label{sec:axpspec} 
\subsubsection{Phase Averaged Spectroscopy} 
\label{sec:axpspec1} 
We use CC mode data to obtain an X-ray spectrum. We 
assume that the CC and TE spectral responses are identical. To test 
this assumption, we compared low count rate data from a particular 
region in both CC and TE mode and found reasonable agreement.  We 
define the source by selecting an interval $\pm 8$ pixels ($\pm \sim 
4\arcsec$) (see also \S \ref{sec:extended}) around the peak 
flux along the collapsed CC mode axis.  The 
background was determined using two adjacent segments 12 pixels wide 
for a total area 1.5 times the source area; the background flux provides 
$\sim2\%$ of the total flux in the source region. 
Source and background spectra and response files were generated 
using the CIAO (v2.1.1) tools {\sl dmextract}, {\sl mkrmf}, and {\sl 
mkarf}. The extraction was performed in pulse invariant (PI) space 
({\em{i.e.}}, after the instrument gains were applied).  The spectrum was 
grouped into bins that contained at least 
25 events. 

\begin{figure*}[r] 
\epsscale{1.0} 
\plotone{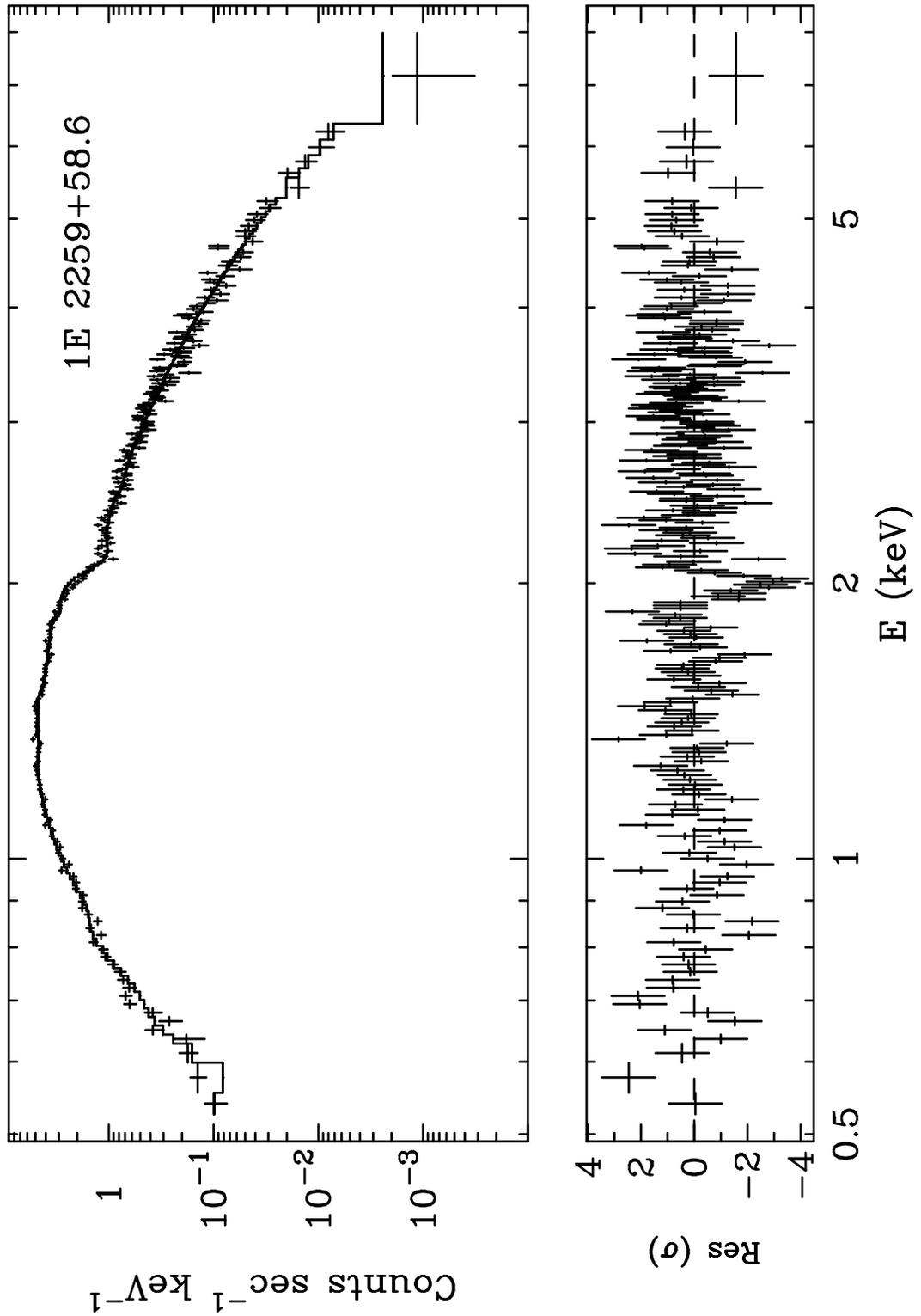}
\caption{The best fit spectral model (PL$+$BB) and residuals 
in units of $\sigma$. The feature at $\sim2.0$~keV appears to be due to a 
small shift of the location of the Ir absorption edge between the response  
and the data. 
} 
\label{fig:avespec} 
\end{figure*}

All spectral fits were limited to the $0.5-7.0$~keV band and use 
XSPECv11.01 \citep{arnaud1996}, the photo-electric absorption 
coefficients of \cite{balucinska1992}, and abundances of 
\cite{anders1989}. We fit the data to a PL+BB model ($\chi^2/d.o.f.= 
316.1/272$); the fit is significantly improved ($\chi^2/d.o.f.= 
262.5/258$) by ignoring the 1.9-2.1 keV band, which contains the 
iridium-edge structure in the telescope response. The best fit PL+BB 
parameters are $\Gamma=3.6\pm{0.1}$, \kt$=0.412\pm{0.006}$~keV, and 
\nh$=0.93\pm{0.03}~\times~10^{22}~{\rm cm}^{-2}$, consistent with the 
same model results of \cite{rho1997}. Assuming a source distance of 
4~kpc, the 2-10 keV unabsorbed flux and X-ray luminosity are 
$2.0\pm{0.2}~\times~10^{-11}$~ergs~s$^{-1}$~cm$^{-2}$ and 
$3.8\pm{0.4}~\times~10^{34}$~ergs~s$^{-1}$ respectively. The PL 
component contributes $50\%$ of the total 2-10 keV unabsorbed flux. 
Figure~\ref{fig:avespec} shows the data, the best-fit model, and the 
residuals. We have also investigated single component PL, BB, and 
bremsstrahlung (BS) models and a combined PL+BS model and found that 
they are all statistically unacceptable at $>99.9, >99.9, >99.7,$ and 
$>96.4\%$ confidence levels, respectively. 


We have searched for spectral lines. Features that appear below $\sim 
1$~keV and in the Ir edge, are likely due to uncertainty in the 
spectral response. To set upper limits, we examined two small 
deviations at 0.7 and 5.0~keV. The addition of a line, modeled by a 
Gaussian with an intrinsic width of 40 eV at 5.0 keV to the PL+BB 
model, results in an F statistic of 3.3; the $90\%$ confidence upper 
limit on the line flux is $8.6 \times 
10^{-13}$~ergs~cm$^{-2}$~s$^{-1}$.  Similarly, including a ``line'' at 
0.7 keV gives an F statistic of 1.3.  We conclude that there are no 
significant line features. 

\subsubsection{Phase Resolved Spectroscopy} 
\label{sec:axpspec2} 
\begin{figure*}[r] 
\epsscale{1.0} 
\plotone{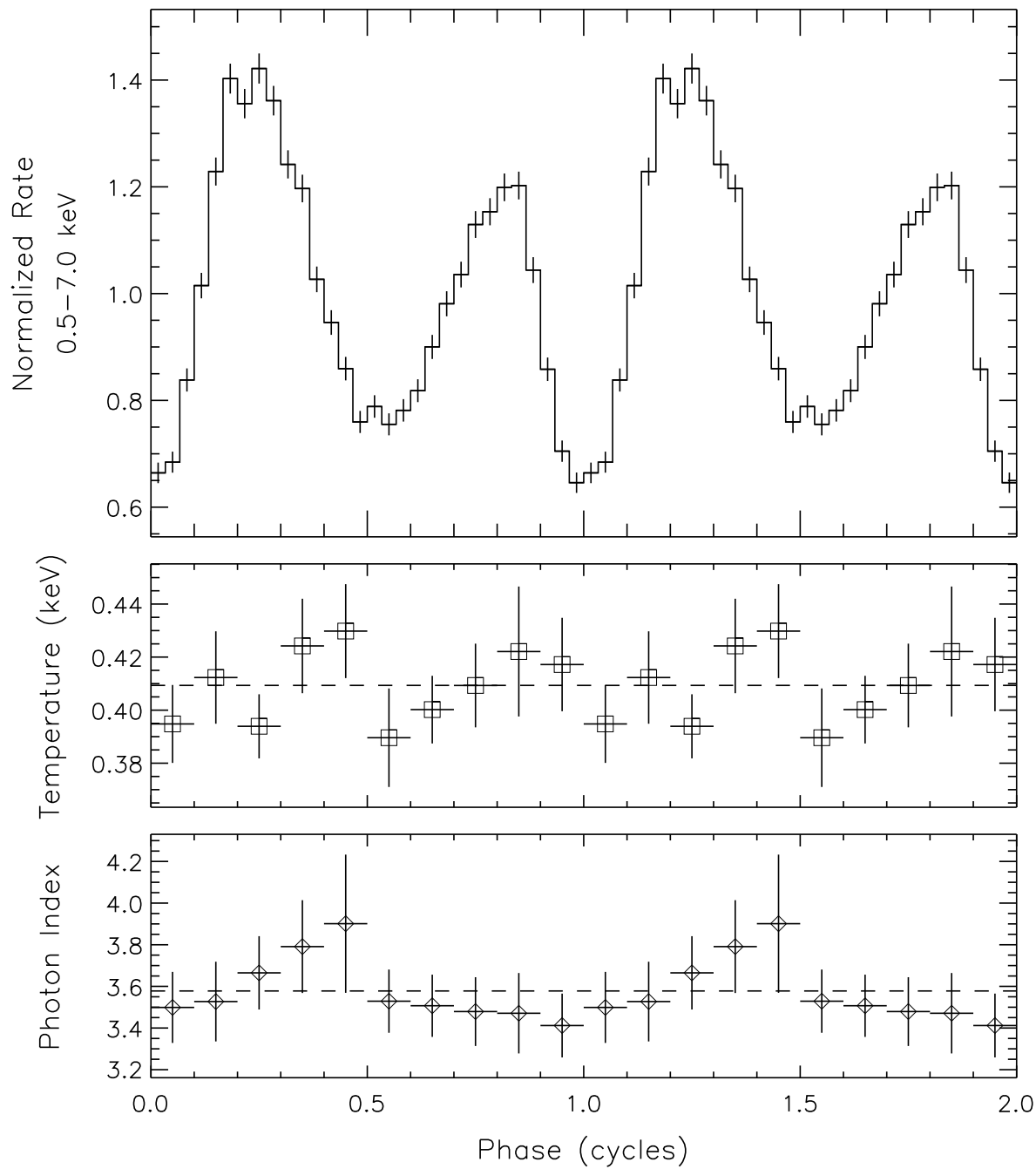}
\caption{Top panel (a): The pulse profile ($0.5-7.0$~keV). 
Middle panel (b): Variation of the blackbody temperature from 
fitting each phase bin with a PL+BB model. The time 
averaged value is denoted by the dashed line. 
Bottom panel(c) : Variation of the PL index. All errors shown denote $1 \sigma$.} 
\label{fig:foldevents} 
\end{figure*}

We fit the data from each of the ten pulse phase bins 
(Figure \ref{fig:foldevents}) with a PL+BB model parameters free to 
obtain their best fit values and with \nh\ constrained to be identical 
at each phase.  The fitting gave a column density of 
$0.93 \pm 0.04~\times~10^{22}~{\rm cm}^{-2}$ 
(and is identical to the time averaged value found in \S \ref{sec:axpspec1}). 
For the remaining fits, \nh\ is constrained to be identical at each phase 
(i.e., linked).  We next fit all phase bins with \kt\ constrained to be 
identical at each phase and allow $\Gamma$ and the 
normalizations to vary.  The decrease in $\chi^2$ was not significant 
indicating the fits were statistically equivalent.  However, there 
were significant variations in $\Gamma$ ($\pm0.2$).  The fits were 
repeated with $\Gamma$ constrained with respect to phase and with \kt\ and 
normalizations allowed to vary.  Again this resulted in a statistically 
equivalent fit but with significant variations of \kt\ ($\pm 0.02$~keV). 
The evolution of spectral parameters is shown as a function of pulse phase in 
Figures~\ref{fig:foldevents}b and \ref{fig:foldevents}c. 
We fit the spectra with \kt\ and $\Gamma$ constrained to be identical at 
each phase with only the normalizations free to vary with phase and find 
that the fits are not statistically equivalent.  These results indicate 
that there is marginal ($\sim 3\sigma$) evidence for evolution in either $\Gamma$, \kt\, or some combination thereof with phase, but we are unable to determine which. Finally, we impose an additional constraint where the ratio of the PL and BB model normalizations (PL$_{\rm Norm}$, BB$_{Norm}$) are constant.  Our fits show no statistical improvement 
in complicating the model by allowing the normalizations to be independent. 
Fit statistics for the models discussed are given in Table \ref{tab:1}. 

\section{Discussion} 
\label{sec:discuss} 
We have observed the anomalous X-ray pulsar, \onee\ with \chan\ and 
determined a new, precise position of the source of 
$\alpha=23^{h}~01^{m}~8^{s}.295$, 
$\delta=+58\arcdeg~52\arcmin~44\arcsec .45$ (J2000) with a $99\%$ 
confidence error radius of $0\arcsec.60$. 
We report in the accompanying paper \citep{hulleman2001} the 
details of the derivation, which led to the identification of a near 
infrared (NIR) counterpart at the \chan\ position.  This is the second 
AXP counterpart to be found in NIR wavelengths \citep{hulleman2000a}. 

We find that of the spectral models we have considered, the PL$+$BB 
one is statistically preferred for the pulsar spectrum. In addition, 
the superb \chan\ spatial resolution allows us to determine that  
there is emission extending up to $100\arcsec$ from the pulsar, but we  
are unable to determine its precise origin - whether from the 
supernova remnant, and/or a dust scattering halo, and/or a plerion. 

\acknowledgments 
We acknowledge support from the following grants: MX-0101 (C.K.), 
NAG5-9350 (P.W.), GO0-1018X (S.P.).  We thank F. Paerels and J. Vink 
for many useful discussions.


\begin{deluxetable}{l l l cc}
\footnotesize
\tablecaption{Phase Resolved Spectral Fit Results for \onee  \label{tab:1}}
\tablewidth{450pt}
\tablehead{
\colhead{} & \colhead{Model Settings\tablenotemark{a}} &\colhead{} & \colhead{}  & \colhead{} \\
\colhead{PL$_{\rm Norm} \propto$ BB$_{Norm}$} & \colhead{$kT_{\rm BB}$ Linked} & \colhead{$\Gamma$ Linked} & \colhead{$\chi^2 / \nu$}  & \colhead{$F-$Statistic\tablenotemark{b}} 
} 
\startdata
Y & Y & Y & $1088.4 / 1050$  & {...}            \\
N & Y & Y & $1072.7 / 1040$  & 1.53 (1.84,2.34) \\ 
N & Y & N & $1049.1 / 1031$  & 2.04 (1.60,1.92) \\
N & N & Y & $1051.1 / 1031$  & 1.92 (1.60,1.92) \\
N & N & N & $1042.9 / 1022$  & 1.59 (1.49,1.74) 
\enddata
\tablenotetext{a}{Model normalizations for each phase bin are free (unconstrained) 
to obtain its best fit value while $N_{\rm H}$ is linked (constrained to be identical 
at each phase) for all model variations.}
\tablenotetext{b}{The numbers in parenthesis correspond to the $F-$statistic needed
to claim that the given model is significantly better at the $95\%$ and $99\%$
confidence levels, respectively.}
\end{deluxetable}


\end{document}